\documentclass{aastex}
\usepackage{emulateapj5}
\usepackage{times,mathptmx}
\slugcomment{Accepted for publication in ApJ Letters}

\shorttitle{The MUNICS Galaxy Stellar Mass Function}
\shortauthors{Drory et al.}

\newcommand\M{\ensuremath{\mathcal{M}}}%
\newcommand\Msun{\ensuremath{\M_\odot}}%
\newcommand\Lsun{\ensuremath{L_\odot}}%
\newcommand\Mlim{\ensuremath{\M_{\mathrm{lim}}}}%
\newcommand\MLK{\ensuremath{\M/L_K}}%
\newcommand\Vmax{\ensuremath{V_{\mathrm{max}}}}%

\begin{document}


\title{The Munich Near-Infrared Cluster Survey (MUNICS) -- Number
  density evolution of massive field galaxies to $z~\sim~1.2$ as
  derived from the $K$-Band Selected Survey}


\author{N.~Drory\altaffilmark{1,2}, R.~Bender\altaffilmark{1},
        J.~Snigula\altaffilmark{1}, G.~Feulner\altaffilmark{1},
        U.~Hopp\altaffilmark{1,3}, C.~Maraston,}
\affil{Universit\"ats-Sternwarte M\"unchen, Scheinerstr. 1,
  D-81679 M\"unchen, Germany}
\email{\{drory,bender,snigula,feulner,hopp,maraston\}@usm.uni-muenchen.de}

\author{G.~J.~Hill\altaffilmark{2},}
\affil{University of Texas at Austin, Austin, Texas 78712}
\email{hill@as.utexas.edu}

\and

\author{C.~Mendes de Oliveira\altaffilmark{1}}
\affil{Instituto Astron\^omico e Geof\'{\i}sico, Av Miguel St\'efano
  4200, 04301-904, S\~ao Paulo, Brazil}
\email{oliveira@iagusp.usp.br}

\altaffiltext{1}{Visiting Astronomer at the German-Spanish
  Astronomical Center, Calar Alto, operated by the Max-Planck-Institut
  f\"ur Astronomie, Heidelberg, jointly with the Spanish National
  Commission for Astronomy.}
\altaffiltext{2}{Visiting astronomer at McDonald Observatory, operated
  by the University of Texas at Austin, and the Hobby - Eberly
  Telescope, operated by McDonald Observatory on behalf of The
  University of Texas at Austin, the Pennsylvania State University,
  Stanford University, Ludwig-Maximilians-Universit\"at M\"unchen, and
  Georg-August-Universit\"at G\"ottingen.}
\altaffiltext{3}{Visiting astronomer at the European Southern Observatory,
  Chile, proposal number N 66.A-0129 and 66.A-0123.}


\begin{abstract}
  We derive the number density evolution of massive field galaxies in
  the redshift range $0.4 < z < 1.2$ using the $K$-band selected field
  galaxy sample from the Munich Near-IR Cluster Survey (MUNICS).  We
  rely on spectroscopically calibrated photometric redshifts to
  determine distances and absolute magnitudes in the rest-frame
  $K$-band. To assign mass-to-light ratios, we use an approach which
  maximizes the stellar mass for any $K$-band luminosity at any
  redshift. We take the mass-to-light ratio, \MLK , of a Simple
  Stellar Population (SSP) which is as old as the universe at the
  galaxy's redshift as a likely upper limit. This is the most extreme
  case of pure luminosity evolution and in a more realistic model
  \MLK\ will probably decrease faster with redshift due to increased
  star formation.  We compute the number density of galaxies more
  massive than $2\times 10^{10} h^{-2} \Msun$, $5\times 10^{10} h^{-2}
  \Msun$, and $1\times 10^{11} h^{-2} \Msun$, finding that the
  integrated stellar mass function is roughly constant for the lowest
  mass limit and that it decreases with redshift by a factor of $\sim
  3$ and by a factor of $\sim 6$ for the two higher mass limits,
  respectively. This finding is in qualitative agreement with models
  of hierarchical galaxy formation, which predict that the number
  density of $\sim M^{*}$ objects is fairly constant while it
  decreases faster for more massive systems over the redshift range
  our data probe.
\end{abstract}



\section{Introduction} \label{s:intro}

Significant progress has been made in understanding galaxy evolution
in recent years due to results of statistically significant and
complete redshift surveys spanning the redshift range $0 < z \lesssim
1$ \citep{CFRS95,ECBHG96,LYCE97,CNOC299} and due to detailed kinematic
and stellar population studies of small samples of galaxies
\citep[e.g.\ ][]{KIFD01,DFKI01,Vogtetal96,BZB96}. On the theoretical
side, semi-analytic models of galaxy formation
\citep{KWG93,SP99,CLBF00} try to reproduce properties of galaxies in
the local universe and predict their evolution with redshift within
the framework of a hierarchical CDM-dominated universe.

The traditional observables used to characterize galaxies are
unsuitable for studying the assembly history of galaxies, one of the
most fundamental predictions of CDM models, since these observables
may be transient. The best observable for this aim is, in principle,
total mass, which is on the other hand very hard to measure. It has
been argued that the best available surrogate accessible to direct
observation is the near-IR $K$-band luminosity of a galaxy which
reflects the mass of the underlying stellar population and is least
sensitive to bursts of star formation and dust extinction
\citep{RR93,KC98a,BE00}. The main uncertainty involved in the
conversion of $K$-band light to mass is due to the age of the
population, amounting to only a factor of two in mass uncertainty for
populations older than $\sim 3$~Gyr.

In this paper we use the $K$-band selected field galaxy sample of the
medium deep, wide-field Munich Near-IR Cluster Survey (MUNICS) to
study the number density evolution of massive galaxies. See
\citet{MUNICS1} for details of the survey (MUNICS1 hereafter).  Data
presented here refer to $\sim 5000$ galaxies for which high quality
VRIJK imaging covering 0.27 square degrees is available (see \S
\ref{s:sample}).  The distances of the galaxies are derived from
spectroscopically calibrated photometric redshifts (\S
\ref{s:photred}) and their masses from rest-frame $K$-band
luminosities (\S \ref{s:integ_mf}).  We discuss the resulting
integrated stellar mass functions at different mass limits and their
evolution with redshift (\S \ref{s:discuss}).

\begin{figure*}
\epsscale{0.65}
\plotone{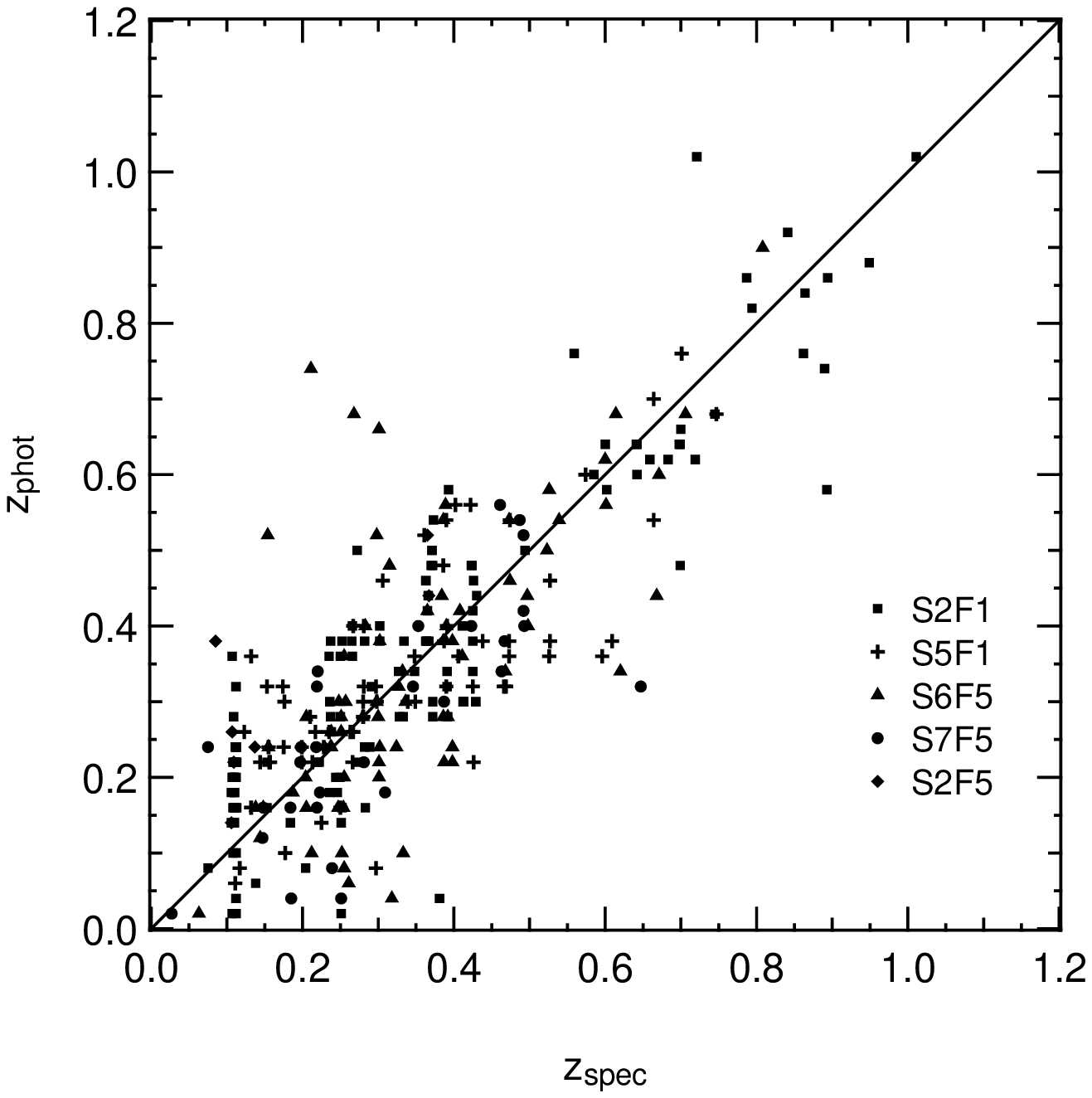}
\plotone{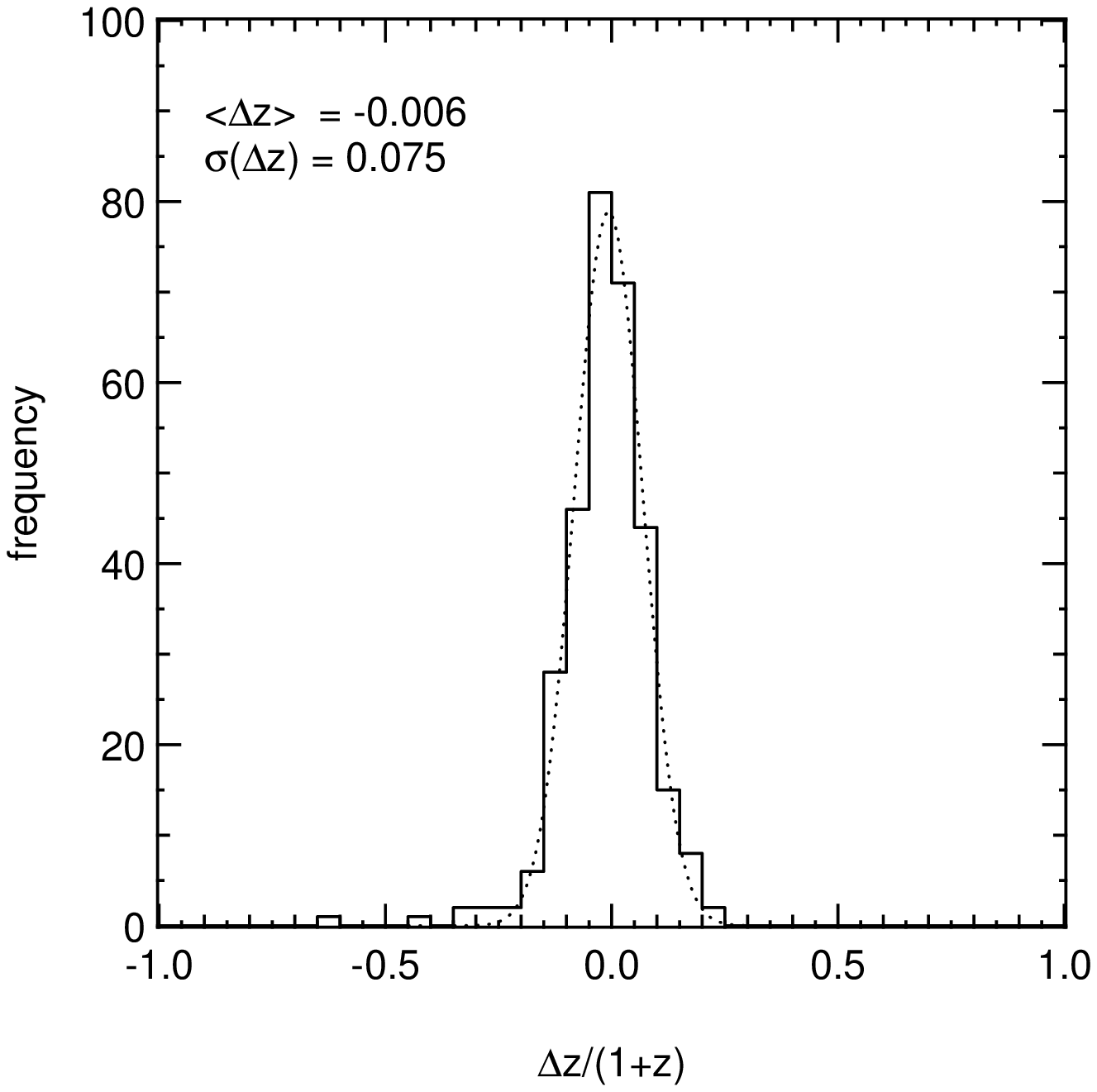}
\plotone{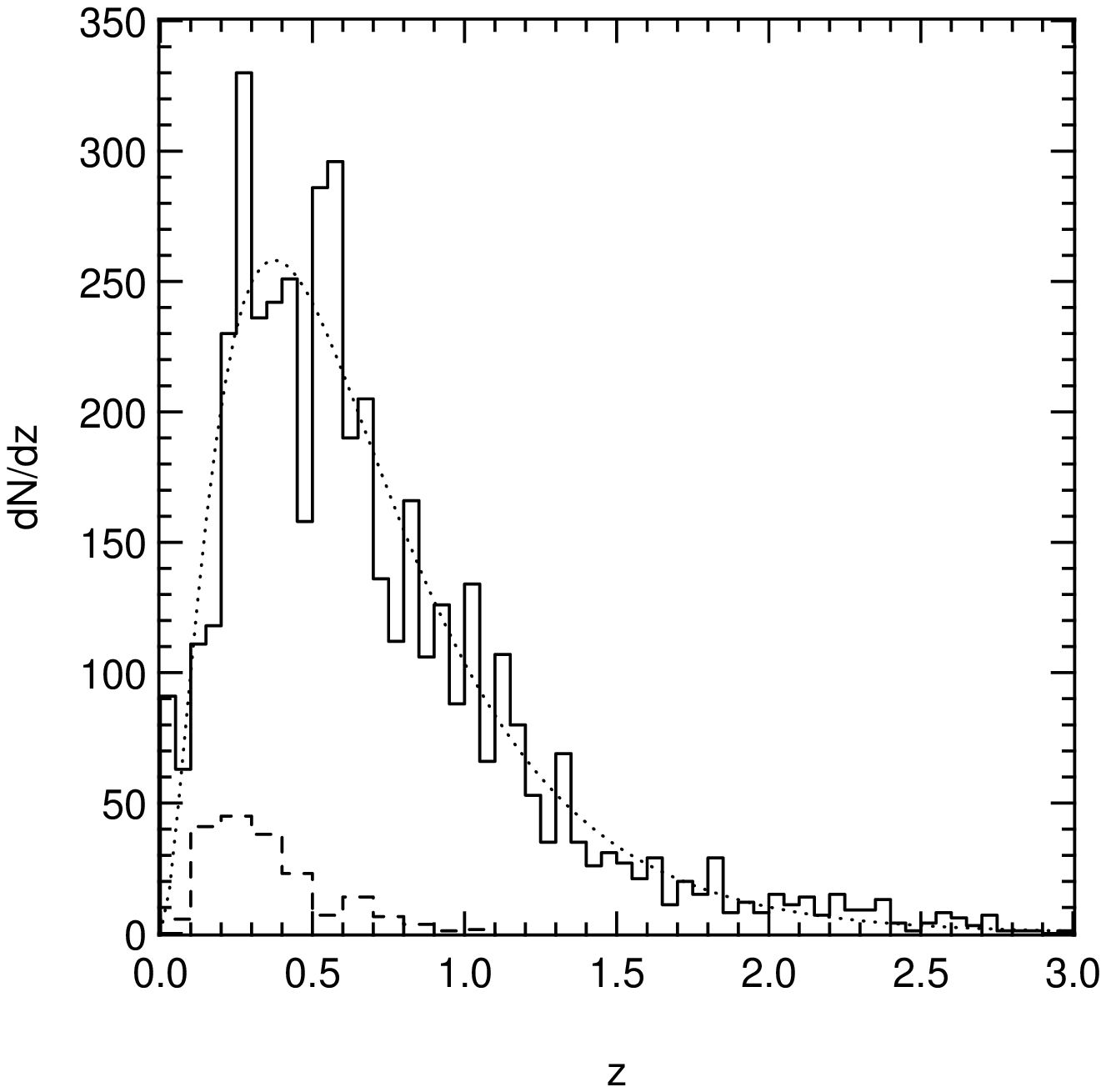}
\caption{\label{f:zz}
  Left panel: Comparison of photometric and spectroscopic redshifts
  for 310 objects in 5 survey patches (different symbols). Middle
  panel: The histogram of the redshift errors. The rms scatter is
  consistent with a Gaussian (dotted line: best-fit Gaussian) of a
  width $\sigma$~=~0.075 and an insignificant mean deviation from the
  unity relation of $<\Delta~z> = -0.006$.  Right panel: The
  distribution of photometric redshifts (solid histogram) and a
  best-fit analytic description (dotted line) as well as the
  distribution of spectroscopic redshifts (dashed line).}
\end{figure*}

We assume $\Omega_M = 0.3$, $\Omega_{\Lambda} = 0.7$ throughout this
paper. We write Hubble's Constant as $H_0 = 100\ h\ \mathrm{km\ 
  s^{-1}\ Mpc^{-1}}$, using $h = 0.60$ unless explicit dependence on
$h$ is given.


\section{The Galaxy Sample} \label{s:sample}

The galaxy sample used here is a subsample of the MUNICS survey,
selected for best photometric homogeneity, good seeing, and similar
depth.  Furthermore, in each of the remaining survey patches, areas
close to the image borders in any passband, areas around bright stars,
and regions suffering from blooming are excluded. The subsample covers
0.27 square degrees in $V$ (23.5), $R$ (23.5), $I$ (22.5), $J$ (21.5),
and $K$ (19.5); the magnitudes are in the Vega system and refer to
50\% completeness for point sources.

Stars were identified following the procedure described in MUNICS1
which relies on the combined information of PSF shape and colors in
the $J\!-\!K$ vs. $V\!-\!I$ plane. The color criteria may also exclude
$z \lesssim 0.25$ compact blue galaxies. Such galaxies are very
unlikely to be present in the $K$-selected sample. Since we restrict
our analysis to $z>0.4$ this is anyway not a problem.

The final catalog covers an area of 997.7 square arc minutes and
contains 5132 galaxies. The fields included in this analysis are S2,
S3f5--8, S5, S6, and S7f5--8. See Table 1 in MUNICS1 for nomenclature
and further information on the survey fields.

\section{Photometric Redshifts} \label{s:photred}

Photometric redshifts were derived using the method presented in
\citet{photred00}. This method is a template matching algorithm rooted
in Bayesian statistics. It closely resembles the method presented by
\citet{Benitez00}.  The templates are derived by fitting stellar
population models of \citet{Maraston98} to combined broad-band energy
distributions of MUNICS galaxies with spectroscopic redshifts. In this
way, representative galaxy templates of mixed stellar populations
(variable age, metallicity, and dust extinction) optimized for the
MUNICS dataset are obtained.

The use of photometric redshifts can introduce systematic errors in
the derived galaxy distances. To investigate these errors, we
performed the following tests: 1) Templates were derived for a
subsample of objects with spectroscopic redshifts and their
suitability for the whole galaxy sample was verified using the
remaining objects with spectroscopy.  2) Using subsets of the template
library, we estimated the errors in the luminosity and mass functions
introduced by a finite (and incomplete) set of templates. These were
included in our error budget in the analysis.  3) Monte-Carlo
simulations were performed to investigate the influence of photometric
redshift errors on the luminosity function (LF) and the galaxy number
densities. The shape of the LF is not significantly biased as long as
the photometric redshifts scatter symmetrically around the true
redshifts \citep[see also][]{SCSK96} and their errors are smaller than
the bin size in $z$ over which the luminosity function is averaged.

Fig.\ \ref{f:zz} shows the comparison of photometric and spectroscopic
redshifts for 310 objects within 5 MUNICS fields.  The typical scatter
in the relative redshift error $\Delta z / (1+z)$ is 0.075. The mean
bias is negligible. The distribution of photometric redshifts peaks
around $z \approx 0.5$. and has a tail extending to $z \approx 3$.

\begin{figure*}
\epsscale{0.7} 
\plotone{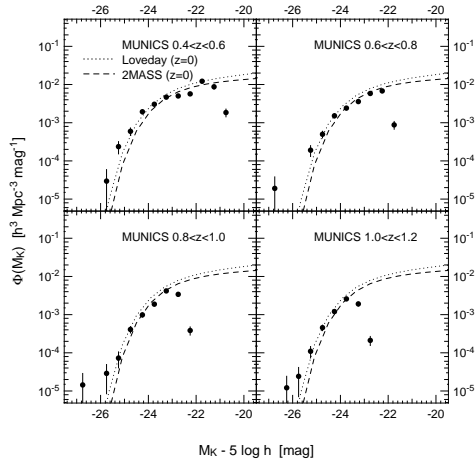}
\caption{\label{f:lf}
  The restframe $K$-band luminosity function derived from the
  MUNICS data in four redshift bins spanning $0.4 < z <1.2$. The
  dotted and the dashed curve are the $z=0$ LF by \citet{Loveday00}
  and \citet{Kochaneketal01}, respectively.}
\end{figure*}

To further strengthen the confidence in the photometric redshifts we
show the restframe $K$-band luminosity function (LF) obtained from
these data in Fig.\ \ref{f:lf}.

The LF shows no significant evolution up to a redshift of 1.2, in
agreement with the finding of \citet{CSHC96} from a much smaller
spectroscopic sample.


\section{The Integrated Stellar Mass Function} \label{s:integ_mf}

The integrated stellar mass function $n(\M>\Mlim)$, the comoving
number density of objects having stellar mass exceeding \Mlim, is
computed using the \Vmax\ formalism to account for the fact that some
fainter galaxies are not visible in the whole survey volume. Each
galaxy in a given redshift bin $[z_l,z_h)$ contributes to the number
density an amount inversely proportional to the volume in which the
galaxy is detectable in the survey:
\begin{equation}
  V_i = d\Omega
  \int_{z_l}^{\mathrm{min}(z_h,z_{max})} \frac{dV}{dz}dz,
\end{equation}
where $dV/dz$ is the comoving volume element, $d\Omega$ is the survey
area, $z_{max}$ is the maximum redshift at which galaxy $i$ having
absolute magnitude $M_{K,i}$ is still detectable given the limiting
apparent magnitude of the survey and the galaxy's SED (the best-fit
SED from the photometric redshift determination).

Additionally, the contribution of each galaxy $i$ is weighted by the
inverse of the detection probability, $P(m_{K,i})$, where we assume
that the detection probability is independent of the galaxy type and
can be approximated by that of point-like sources. We use only objects
with $P(m_{K,i}) > 0.75$, such that this correction is always small.
We have checked that this correction does not bias our results by
comparing to what we get for higher completeness limits.

To compute the stellar mass of a galaxy, we use an approach which
maximizes the stellar mass for any $K$-band luminosity at any
redshift. Noting that \MLK\ is a monotonically rising function of age
for Simple Stellar Populations (SSPs), we find that the likely upper
limit for \MLK\ is the mass-to-light ratio of a SSP which is as old as
the universe at the galaxy's redshift. This is the most extreme case
of passive luminosity evolution (PLE) one can adopt. It corresponds to
a situation where all massive galaxies would be of either elliptical,
S0, or Sa type. We take the mass-to-light ratios from the SSP models
published by \citet{Maraston98}, using a Salpeter IMF. Similar
dependencies on age are obtained from the models of \citet{Worth94}
and \citet{BC93} although the absolute values of \MLK\ vary somewhat.
The \MLK\ which we obtain with our cosmological parameters at
$z~\sim~0.5$ are approximately consistent with \MLK\ of local galaxies
\citep{BD01}. To obtain a more realistic estimate of \MLK, we used our
VRIJK color information and the photometric redshift to fit the age
and SFR of each galaxy using a grid of composite stellar populations
with exponential star formation timescales ranging from 1 to 8~Gyr. We
obtain the same slope in the evolution of the average \MLK\ with
redshift as the one obtained under the PLE assumption. This modelling
will be described in detail in a forthcoming paper.

The mass in solar units is given by
\begin{equation}
  \M = \left( \frac{\M}{L_K} \right) 10^{-0.4(M_K - M_{\odot,K})},
\end{equation}
using $M_{\odot,K} = 3.33$.

The number density in each redshift bin, $n(\M>\Mlim,z)$, is finally
computed by summing over all galaxies in the bin whose stellar mass is
exceeding \Mlim,
\begin{equation}
  n(\M>\Mlim,z) = \sum_i \frac{1}{V_i P(m_{K,i})}.
\end{equation}

\begin{figure*}
\epsscale{2}
\plotone{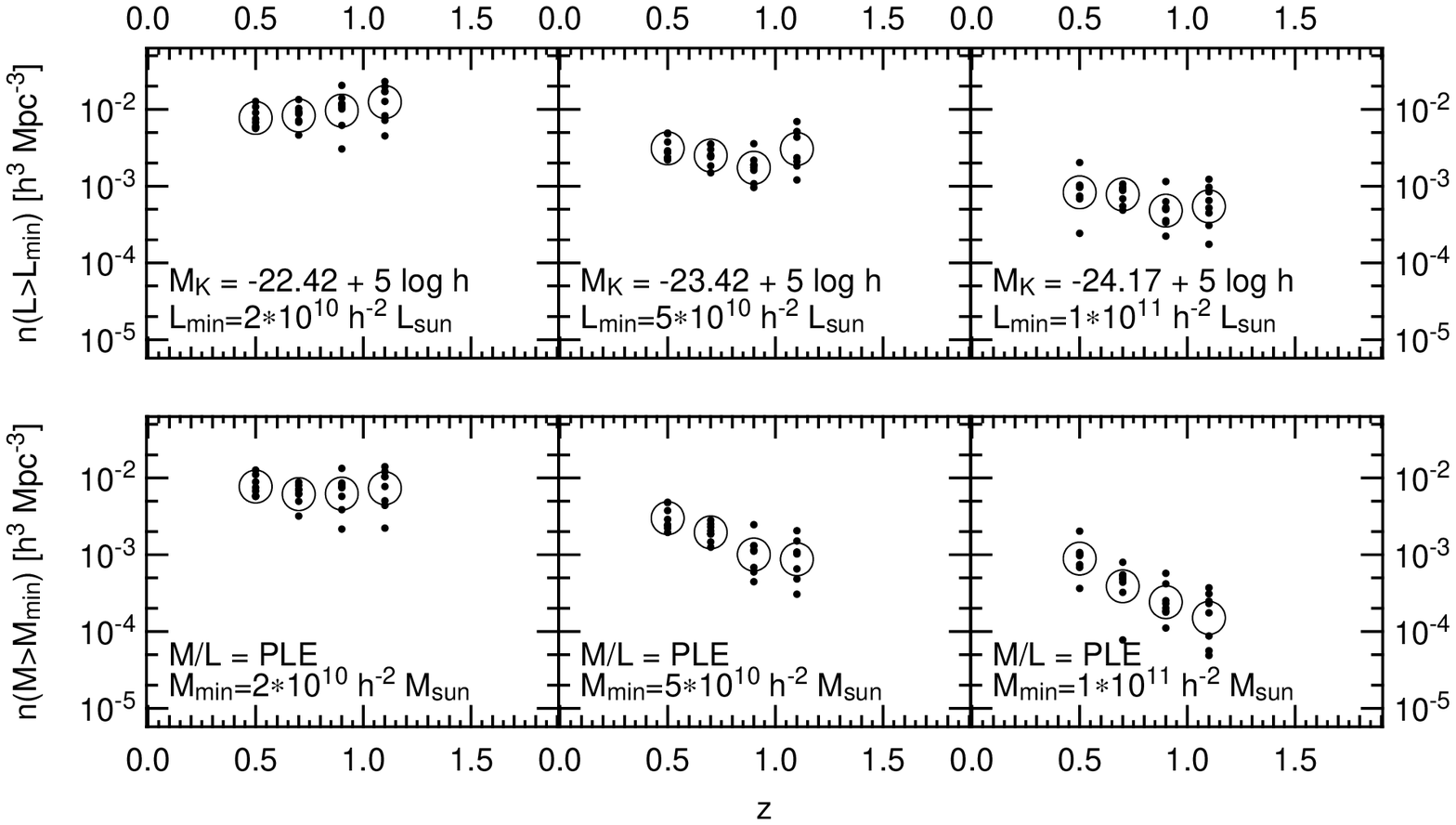}
\caption{\label{f:integ_mf}
  Comoving number density of objects having rest frame $K$-band
  luminosities exceeding $-22.42 + 5\log h$ ($2\times 10^{10} h^{-2}
  \Lsun$), $-23.42 + 5\log h$ ($2\times 10^{10} h^{-2} \Lsun$), and
  $-24.17 + 5\log h$ ($2\times 10^{10} h^{-2} \Lsun$) (upper panels)
  and comoving number density of objects having stellar masses
  exceeding $\Mlim = 2\times 10^{10} h^{-2} \Msun$, $\Mlim = 5\times
  10^{10} h^{-2} \Msun$, and $\Mlim = 1\times 10^{11} h^{-2} \Msun$
  (integrated stellar mass functions; lower panels).  Mass to light
  ratios are assigned to maximize the stellar mass at a given
  luminosity (see text), and thus are likely upper limits.  The solid
  points denote the values measured separately in each survey field,
  the open circles denote the mean values over the whole survey area.
  The size of the open circles is chosen to represent our estimate of
  the total uncertainty in the mean values.}
\end{figure*}

The resulting integrated mass functions for $\Mlim = 2\times 10^{10}
h^{-2} \Msun$, $\Mlim = 5\times 10^{10} h^{-2} \Msun$, and $\Mlim =
1\times 10^{11} h^{-2} \Msun$ are shown in Fig.\ \ref{f:integ_mf}
along with the integrated luminosity functions for comparison. The
mean values of \MLK\ in the maximum PLE model in the four redshift
bins are $0.99$, $0.83$, $0.73$, and $0.65$, as computed from the
look-back time in our cosmology. With these mean values the mass limits
correspond to absolute $K$-band magnitudes of $-22.43$, $-22.63$,
$-22.77$, and $-22.90$, respectively, for $\Mlim = 2\times 10^{10}
h^{-2} \Msun$.  For $\Mlim = 5\times 10^{10} h^{-2} \Msun$ the numbers
are $-23.42$, $-23.62$, $-23.76$, and $-23.89$. Finally, for $\Mlim =
1\times 10^{11} h^{-2} \Msun$ we have $-24.18$, $-24.38$, $-24.51$,
and $-24.64$ (magnitudes with respect to $h = 1$). 



\section{Discussion} \label{s:discuss}

Fig.\ \ref{f:integ_mf} compares the evolution of the integrated
luminosity to the integrated mass.  It is evident that the number
density of {\em luminous} $K$-band selected galaxies does not evolve
significantly (given our uncertainties) to $z = 1.2$. However, because
of the inevitable evolution of the mass-to-light ratio with $z$, the
number density of {\em massive} systems does change.  Transforming
luminosities into masses with our maximum PLE scheme yields a roughly
constant number density for our lowest mass limit, $2\times 10^{10}
h^{-2} \Msun$, and a decrease of the number density with redshift by a
factor of $\sim 3$ for a mass limit of $5\times 10^{10} h^{-2} \Msun$,
and by a factor $\sim 6$ for objects more massive than $1\times
10^{11} h^{-2} \Msun$.  As the true \MLK\ at high redshift will most
likely be lower than in our maximum PLE model, the true number
densities are likely to decrease more rapidly with redshift.

The steepening of the curves with increasing limiting mass in the
maximum PLE case (despite them all having the same mass-to-light
ratios at any given redshift) is due to the invariance of the LF with
redshift and its steepness at the bright end. At increasing limiting
mass, one is moving down the steepening bright end of the LF, so that
the same change in the mass-to-light ratio yields a higher change in
the number density.

The main uncertainty in this conclusion is still the field to field
variation, in spite of the relatively large area surveyed, followed by
the choice of SED templates used in the photometric redshift code (see
above). The size of the open symbols in Fig.\ \ref{f:integ_mf}
represents our estimate of the total uncertainty of the mean values.
If we assume a Gould IMF \citep{GFB98} instead of a Salpeter IMF, the
evolving \MLK\ curve becomes lower in its normalization as the
mass-to-light ratio becomes smaller due to the reduced number of
low-mass stars. The slope does not change significantly.

The observed density evolution as a function of mass is qualitatively
consistent with the expectation from hierarchical galaxy formation
models. Most rapid evolution is predicted for the number density of
the most massive galaxies while the number density of $L^*$-galaxies
should evolve much less.  E.g.\ \citet{BCFL98} predict that the number
density of galaxies of a stellar mass of $10^{10} h^{-1} \Msun$
decreases by a factor of $\sim 3.1$ over redshift range $0.4 < z <
1.2$ (for the cosmological parameters as used here).  Though this
agreement is encouraging, both more elaborated models and improved
sets of data are required.  The latter can be obtained by larger and
deeper samples, and more realistic estimates of \MLK.


\acknowledgments

We would like to thank the Calar Alto staff for their long-standing
support during many observing runs over the last five years.  This
work was partly supported by the Deutsche Forschungsgemeinschaft,
grant SFB 375 ``Astroteilchenphysik'' and the German Federal Ministry
of Education and Research (BMBF), grant 05 AV9WM1/2.  ND would like to
thank S.\ White and G.\ Kauffmann for helpful discussions. GJH
acknowledges support by the Texas Advanced Research Program under
Grant No. 009658-0710-1999. CMdO acknowledges support from FAPESP
(Funda\c c\~ao de Amparo a Pesquisa do Estado de S\~ao Paulo) and the
Alexander von Humboldt Foundation.


\bibliography{apjmnemonic,literature} \bibliographystyle{apj}

\end{document}